\documentclass[conference]{IEEEtran}
\usepackage[T1]{fontenc}
\usepackage{stackengine}
\usepackage{cite}
\usepackage{amsmath,amssymb,amsfonts, array}
\usepackage[linesnumbered,ruled,vlined]{algorithm2e}
\usepackage{graphicx}
\usepackage{textcomp}
\usepackage{xcolor}
\usepackage{mathtools, stackengine, mathrsfs, setspace, bm, multirow, tikz, pgfplots, adjustbox, calc, url, xtab}
\usepackage[colorlinks=true, citecolor=blue, linkcolor=black!20!red]{hyperref}
\usepackage[caption=false,font=footnotesize]{subfig}
\usepackage{xparse}
\usetikzlibrary{arrows,automata}

\allowdisplaybreaks

\begin{document}

%\bstctlcite{IEEEexample:BSTcontrol}
\title{An Insurance Paradigm for Improving Power System Resilience 
		via Distributed Investment 
}

\author{\IEEEauthorblockN{Farhad Billimoria\IEEEauthorrefmark{2}\IEEEauthorrefmark{1}, 
Filiberto Fele\IEEEauthorrefmark{2},
Iacopo Savelli\IEEEauthorrefmark{4}, 
Thomas Morstyn\IEEEauthorrefmark{3}, 
Malcolm McCulloch\IEEEauthorrefmark{2}}
\IEEEauthorblockA{\IEEEauthorrefmark{2}Energy \& Power Group, Department of Engineering Science, University of Oxford}
\IEEEauthorblockA{\IEEEauthorrefmark{3}School of Engineering, University of Edinburgh}
\IEEEauthorblockA{\IEEEauthorrefmark{4}Centre for Research on Geography, Resources, Environment, Energy \& Networks, Bocconi University}
\IEEEauthorblockA{\IEEEauthorrefmark{1}Corresponding author: farhad.billimoria@wolfson.ox.ac.uk}
}

\maketitle

\begin{abstract}
Extreme events, exacerbated by climate change, pose significant risks to the energy system and its consumers.  However there are natural limits to the degree of protection that can be delivered from a centralised market architecture. Distributed energy resources provide resilience to the energy system, but their value remains inadequately recognized by  regulatory frameworks. We propose an insurance framework to align residual outage risk exposure with locational incentives for distributed investment. We demonstrate that leveraging this framework in large-scale electricity systems could improve consumer welfare outcomes in the face of growing risks from extreme events via investment in distributed energy. 
\end{abstract}
\section{Introduction}\label{sec1}
This paper addresses the issue of incentive frameworks for decentralised resilience investments by proposing an electricity interruption insurance scheme to price residual outage risk. In doing so it presents a defensible rationale for efficient investment in resilient distributed energy resources (DER).\par
The nature of risks facing the electricity system is changing. The impetus for sectoral decarbonisation is expected to drive order of magnitude increases in generation supply from variable renewable energy (VRE), to be managed with the rolloff of an ageing and increasingly unreliable thermal fleet \cite{Simshauser2018,Billimoria2019}. With climate change already occurring, the frequency and severity of extreme weather events are expected to magnify  with particularly significant impacts on centralised grid architectures \cite{Bennett2021}. \par
While wholesale energy markets can in theory ensure a reliable system \cite{Schweppe1988} a range of recent works identify incompleteness in liberalized market architectures that can leave systems and communities vulnerable to extreme events \cite{Wolak2021a,Joskow2007,Billimoria2022}. Administrative contracting too can distort the fuel mix  towards resources that are particularly vulnerable to weather extremes \cite{Mays2019,Mays2021b}. Furthermore, to no flaw of market design, extreme events can island particular regions leaving communities disrupted and at risk for sustained periods. The recognition that, despite best efforts, wholesale market design inevitably leaves residual outage exposure for consumers leads to a view by some that promising complete protection from wholesale market frameworks is at best inordinately costly, and at worst illusory \cite{Hogan2022}. Yet there is a concomitant acknowledgement that leaving open such vulnerability may also be undesirable, particularly given a changing climate \cite{NationalAcademiesofSciences2017} and the inequitable impacts of outages from extreme events \cite{Meerow2019,Lievanos2017,Ulak2018,Dargin2020,Markhvida2020}. \par
Decentralised technologies offer the technical potential for improved resilience to extreme events. In particular, solar PV, storage and other DER (EVs, smart home etc), can be configured to act as micro-, nano-, and pico- grids during emergencies providing power at community and consumer levels when centralised system architectures fail \cite{Baik2018a,Bennett2021,Hotaling2021,Gorman2022}. An economic framework that appropriately values the resilience benefits of DER technologies could catalyse investment that enables the realization of this technical potential. In this paper we are primarily interested in the concept of reliability insurance as an economic  framework for valuing resilience via distributed investment.

\subsection{Related Work and Contributions}

The concept of reliability insurance in electricity market design was introduced in \cite{Hung-PoChaoandRobertWilsonSource2016,Chao1988,Oren1990} as a means of pricing priority service. The central precept of this line of work involves an insurance contract between an energy consumer and an insurance agency which provides economic compensation for electricity interruption in exchange for an upfront premium. The application of insurance schemes to incentives for backup generation was investigated in \cite{Doucet1991},\cite{Fuentes2020} and \cite{Billimoria2022}. A key result of \cite{Doucet1991} was that, under full insurance, an individual's economic incentives to install onsite backup generation to minimise premia will supplant the utility's incentive to mitigate compensation liabilities. \cite{Fuentes2020} uses an agent-based model that confirms that insurance contracts converge to theoretical optima under bounded perception of risks and losses. The authors in \cite{Billimoria2022} establish that a compensatory insurance scheme can improve consumer outcomes in the presence of reliability externalities. More recently, \cite{Shmuel2022} establishes that priority service Pareto dominates both ex-ante time-of-use pricing and integrated resource planning under supply uncertainty. However, all of these works adopted a simplified copper-plate network model and a generalised definition of reserves. \par
Our paper uses a more detailed model of the network and the market design to understand the regional and temporal aspects of reliability insurance as an incentive framework for distributed investment. Specifically the contributions and novel aspects of this paper are as follows:
\begin{enumerate}
\item we develop a locational model of reliability insurance that differentiates risk on a regional level, recognizing remoteness and weak network connectivity. This can play an important role in scalable management of extreme risks;
\item we formulate a multi-agent model of the electricity system to test the effect of the insurance mechanism on three different market designs. The model reflects the spatial topology of the grid with the objective of providing insight on participant behaviour, the nature of interactions between participant and design, and system reliability and resiliency impacts;
\item we propose two investment incentive frameworks for resilient DER - direct investment and subsidization. It is shown that while subsization can leverage consumer self-insurance benefits, takeup depends upon risk aversion, which is non-transparent to regulators.\end{enumerate}
%\item finally to address the %distributional equity concerns of %remote regions bearing a higher %insurance burden, we present a %re-allocation method that enables an %agency to tradeoff efficiency and %fairness in premium setting.
The structure of this paper is as follows. Section \ref{:sec architect} presents the technical and market architecture associated with the insurance scheme for resilient distributed energy resource investment. In Section \ref{:sec methods} we present the multi-agent model of the market and the program formulation for insurance-based decision making. Section \ref{:sec numerical} applies the model to a numerical case study, followed in Section \ref{sec: conc} by policy implications and conclusions.\par
\section{System and Risk Architecture} \label{:sec architect}
\begin{figure*}[thpb]
	\centering
	\includegraphics[height=0.8\columnwidth]{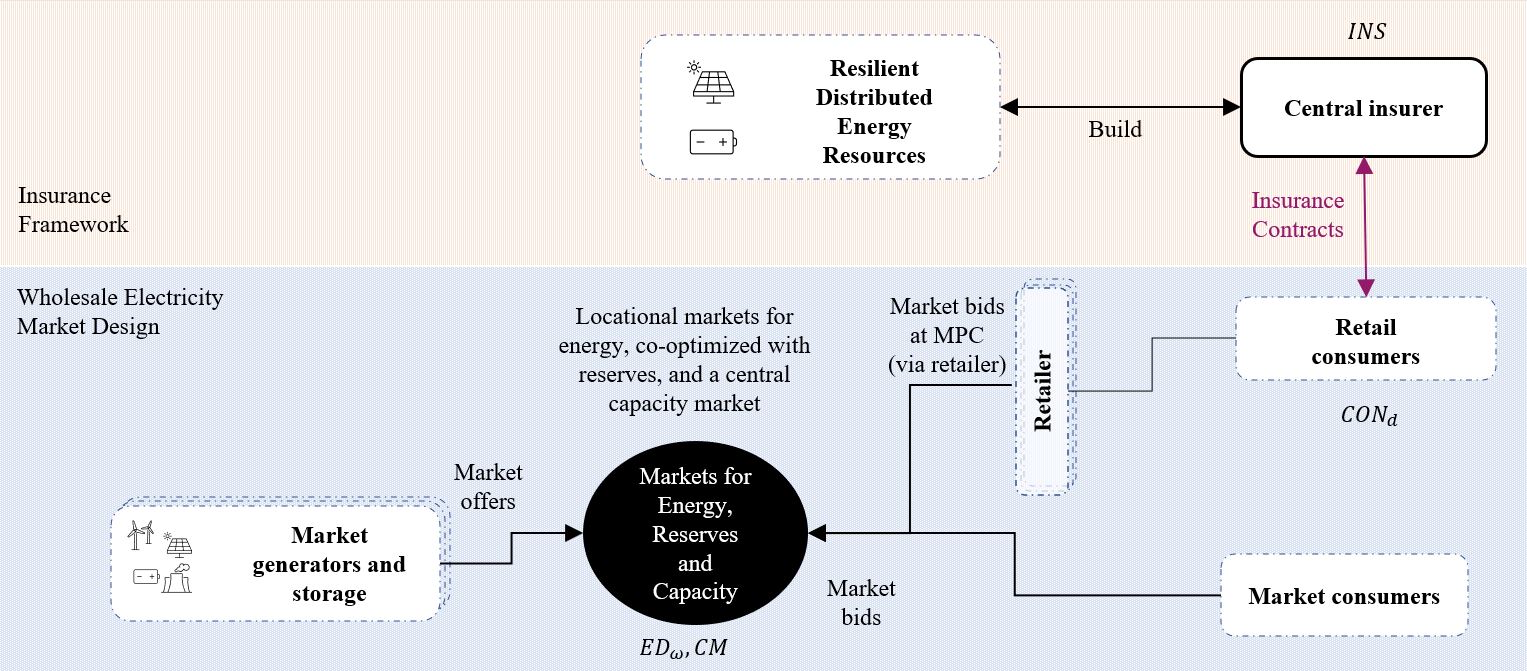}
	%	\begin{tabular}{cc}     
	%	\end{tabular}
	\caption{Schematic of the market architecture incorporating a wholesale electricity market design with centrally cleared spot markets for energy, reserves and capacity; combined with an insurance scheme for electricity sector resilience.}
	\label{fig:architect}
\end{figure*}

Our approach in this paper involves the imposition of an insurance scheme to insure consumers for long-duration outages. Figure \ref{fig:architect} shows a high level schematic of the scheme. \par
Two components make up the market architecture. First, the wholesale market design, which comprises a spot market combined with additional resource adequacy mechanisms. In this paper, we consider a locational spot market for electricity, that is optionally augmented with an operating reserve demand curve (ORDC) and a capacity mechanism \cite{Cramton2017}.\par
The second component of the architecture is an insurance mechanism for system resiliency. The insurer offers electricity interruption insurance to consumers. In exchange for an upfront insurance premium, insurance provides consumers with financial compensation in the event that load is interrupted, in the form of a payment (represented in \$ per MWh) linked to the value of the particular source of consumption, consistent with optimal contract selection \cite{zhao2022}. The insurer manages the tail risks by setting premiums and reserving capital against severe losses. In addition, to reduce its risk of compensation liability exposure the insurer can also offset risk through investment in resilient distributed energy resources (RDER). The incentive to invest is dependent upon the extent to which the investment mitigates service interruption, aligning interests between the insurer and consumers. We consider two forms of investment: (i) direct investment, where the insurer bears the full investment cost of RDER; and (ii) subsidy, where the insurer subsidises the investment cost of RDER for consumers.\par
We adopt a RDER system architecture that incorporates (i) a distributed solar system and (ii) a battery energy storage system that is connected to the central grid and enabled for islanded operation if the grid connection is interrupted. This represents one potential setup of DER that could aid in improving resilience to extreme events.\footnote{Other options include feeder and substation level configurations (see for example \cite{Baik2018b}). Centralized network resilience enhancement could also be considered and co-optimized with distributed investment options, though we keep this out of scope for this paper to keep the formulation tractable.}.\par
\section{Methods} \label{:sec methods}
To illustrate the economic rationale for the proposal we develop an agent-based model of investment in the electricity market and via the associated insurance scheme. \par
Subsection \ref{:sec wholesale_methods} presents the decision formulation for agents in the wholesale electricity market. Subsection \ref{:sec ins_methods} presents the decision making formulation for the insurer and consumers under an insurance overlay. Subsection \ref{:sec market_eq_alg} presents an algorithm to find an equilibrium among participants in the market and insurance scheme.\par
\subsection{Investment decision-making in wholesale markets} \label{:sec wholesale_methods}
In this subsection we present the mathematical formulation of the multi-agent model of the electricity market.\par  
Each generation or storage resource adopts a two-stage decision making process. Investment decisions are made in the first stage based on outcomes in the second stage. The second stage represents the economic dispatch of energy and operating reserves and clearing of the capacity mechanism. \par
Four aspects of uncertainty are modelled (locational demand, resource availability, network availability and inflows into hydro storage) reflected in annual scenarios ($\omega \in \Omega$).\par
\subsubsection{Economic dispatch formulation}
The electricity spot market ${ED}_\omega$, as formulated below, is represented by a centrally cleared bid-based economic dispatch for energy and operating reserves.\par
The set of resources $r \in \mathcal{R}$ comprise generation $\mathcal{G}$, storage $\mathcal{S}$ and hydro $\mathcal{H}$ units $(\mathcal{R}=\mathcal{G} \cup \mathcal{S} \cup \mathcal{H})$, based on capacity investment decisions in the first stage.\par
For ease of notation, any decision variables and parameters that vary over time are represented by a \textbf{bold} vector. For example, we define the vector of energy dispatched from a resource $r$ over time as $\mathbf{p^{G}_{r\omega}} \coloneqq [p^{G}_{r1\omega},...,p^{G}_{rt\omega},...,p^{G}_{r \mathcal{T} \omega}]$ where $p^{G}_{rt\omega}$ denotes the energy dispatched by resource $r$ in scenario $\omega$, time period $t \in \mathcal{T}$. Other vectors are defined similarly. All mathematical notation is stated in Supplementary Information Section I.\par
For storage resources, energy dispatch is separated into charge $p^{G+}_{st\omega}$ and discharge $p^{G-}_{st\omega}$ with total energy generation defined as the difference between the two  $p^{G}_{st\omega}=p^{G-}_{st\omega}-p^{G+}_{st\omega}$. Total reserve dispatch is defined as the sum of reserve delivery from $p^{R}_{st\omega}=p^{R-}_{st\omega}+p^{R+}_{st\omega}$. This avoids the need for integer variables in the formulation.
\par 
We only model upward reserve here but the model can be readily extended to incorporate additional reserve markets.\par 
\begin{multline}
	\min ED_\omega = \sum_{r \in \mathcal{R}} \mathbf{C^{vc}_{r \omega}}.\mathbf{p^{G}_{r\omega}} + \sum_{d \in \mathcal{D}} \mathbf{C^{sh}_{d \omega}}.\mathbf{p^{sh}_{d\omega}} \\ + \sum_{r \in \mathcal{R}} \mathbf{C^{R}_r}.\mathbf{p^{R}_{r\omega}} + \sum_{i \in \mathcal{I}}  \mathbf{C^{rsh}_i} \mathbf{p^{rsh}_{i\omega}}
	.\label{LL_lower_ED} 	
\end{multline}
subject to: 
\begin{multline}
	\sum_{d \in \mathcal{D}^n} (\overline{\mathbf{P}}^D_{d\omega} - \mathbf{p^{sh}_{d\omega}}) + \sum_{m \in \mathcal{L}^n} B_{nm} (\mathbf{\theta_{n \omega}}-\mathbf{\theta_{m \omega}}) = \sum_{r \in \mathcal{R}^n} \mathbf{p^{R}_{r\omega}}, \\ \omega \in \Omega,n \in \mathcal{N} [\mathbf{\lambda^E_{\omega n}}] \label{edisp}
\end{multline}
\begin{IEEEeqnarray}{l'l'l}
	\mathbf{p^{sh}_{d\omega}} \leq \overline{\mathbf{P}}^D_{d\omega} & \forall d \in \mathcal{D},\omega \in \Omega &  \label{short_min1} \\
	\mathbf{p^{G}_{r\omega}} +\mathbf{p^{R}_{r\omega}} \leq  \overline{{P}_{r}} \mathbf{{A}^G_{r\omega}}u_r  & \forall r \in \mathcal{G} \cup \mathcal{H},\omega \in \Omega &  \label{max_gen1}\\
	\mathbf{p^{G+}_{r\omega}} -\mathbf{p^{R+}_{r\omega}} \leq  \overline{{P}_{r}}\mathbf{{A}^G_{r\omega}} u_r  & \ \forall r \in \mathcal{S},\omega \in \Omega &  \label{max_stor_ch}\\
	\mathbf{p^{G-}_{r\omega}} +\mathbf{p^{R-}_{r\omega}} \leq  \overline{{P}_{r}} \mathbf{A^G_{r\omega}} u_r  & \forall r \in \mathcal{S},\omega \in \Omega &  \label{max_stor_dis}
\end{IEEEeqnarray}
\begin{multline}
    - F_{nm} \mathbf{A^L_{nm,\omega}} \leq B_{nm}( \mathbf{\theta_{\omega n}}-\mathbf{\theta_{\omega m}}) \leq F_{nm} \mathbf{A^L_{nm,\omega}} \\ \forall n \forall m \in \mathcal{L}^n,\ \omega \in \Omega  \label{trans}
\end{multline}
\begin{align}
    S_{rt\omega} = S_{r,t-1,\omega} + q^{+}_{r} p^{G+}_{rt\omega} - \frac{1}{q^{-}_{r}} p^{G-}_{rt\omega}, \ \forall r \in \mathcal{S} ,t \in \mathcal{T},\omega \in \Omega  \label{stor_temp1} \\
	S_{rt\omega} = S_{r,t-1,\omega} + i^{G+}_{rt\omega} - \frac{1}{q^{-}_{r}} p^{G-}_{rt\omega}, \ \forall r \in \mathcal{H} ,t \in \mathcal{T},\omega \in \Omega \label{stor_temp2}
\end{align}
\begin{align}
    \mathbf{S_{r\omega}} \leq {\overline{{P}_{r}} u_r}{e_{r}}, \ \forall r \in \mathcal{S} \cup \mathcal{H},\omega \in \Omega  \label{stor_soc}\\
	\sum_{r \in \mathcal{R}} \mathbf{p^{R}_{r\omega}} +\sum_{i \in \mathcal{I}} \mathbf{p^{rsh}_{i\omega}} \geq \overline{R^{req}}, \ \forall r \in \mathcal{R},\omega \in \Omega [\mathbf{\lambda^{R}_{\omega}}]\label{res_req} 
\end{align}
\begin{IEEEeqnarray}{l}
	\mathbf{p^{rsh}_{i\omega}} \leq  R^{req}_{i},\ \forall i \in \mathcal{I}, \omega \in \Omega \label{max_rsh1}\\
	\mathbf{\theta_{\omega 1}} = 0,\omega \in \Omega \label{ref_node}\\
	\mathbf{p^{G+}_{r\omega}} \geq 0, \mathbf{p^{G-}_{r\omega}} \geq 0,\mathbf{p^{R+}_{r\omega}} \geq 0, \mathbf{p^{R-}_{r\omega}} \geq 0,\mathbf{S_{r\omega}} \geq 0 \label{min_gen1}
\end{IEEEeqnarray} \par
The objective function (\ref{LL_lower_ED}) minimises total costs. The first term represents energy generation costs as the cross product of energy dispatched ($\boldsymbol{p}^{G}_\omega$) and unit variable costs ($\boldsymbol{C^{vc}}_{r \omega}$). The second term represents costs of unserved demand, the cross product of unserved demand ($\boldsymbol{p}^{sh}_{d \omega}$) and the value of lost load ($\mathbf{C^{sh}_{d \omega}}$). The third term is the cost of operating reserves, as the product of operating reserve dispatch ($\mathbf{p^{R}_{r\omega}}$) and unit reserve costs ($\mathbf{C^{R}_r}$). The final term is the cost of unmet reserves, as the product of unmet reserves ($\mathbf{p^{rsh}_{i\omega}}$) and system penalty costs of unmet reserves ($\mathbf{C^{rsh}_i}$).\par
Nodal power balance is defined in equation (\ref{edisp}), with the constraint dual ($\mathbf{\lambda^E_{\omega n}}$) representing the locational marginal price of energy. Equation  (\ref{short_min1}) ensures that unserved demand is below actual nodal demand. Equations (\ref{max_gen1}),(\ref{max_stor_ch}) and (\ref{max_stor_dis}) ensure the energy and reserve dispatch are below the deliverable capacity, represented as the product of resource capacity $\overline{{P}_{r}}$, temporal availability $\mathbf{{A}^G_{r\omega}}$ and the (binary) build status of the resource $u_r$. Equation (\ref{trans}) enforces transmission DC flow limits. Equations (\ref{stor_temp1}) and (\ref{stor_temp2}) define state-of-charge (SoC) for storage and hydro, with  hydro SoC dependent upon rain flow over time ($i^{G+}_{rt\omega}$). These two equations cannot be defined in vector form as they depend upon the charge status in the prior interval. SoC is constrained to have the same value at start and end of each representative period \cite{Jenkins2017}. Techncial limits on SoC are enforced in  (\ref{stor_soc}). Equation (\ref{res_req}) determines reserve balance with the dual ($\mathbf{\lambda^{R}_{\omega}}$) representing the system marginal reserve price. Equation (\ref{max_rsh1}) limits reserve shortages based on an segmented operating reserve demand curve (ORDC) \cite{Aderhold2011}. Equations (\ref{ref_node})-(\ref{min_gen1}) set reference phase angles and non-negativity constraints. \par
\subsubsection{Capacity mechanism formulation}
The formulation for the capacity mechanism ${CM}$ envisions a central auction for resource capacity cleared against an administratively-determined demand curve. The capacity mechanism provides an additional source of revenue to resources based on the marginal price of the capacity auction and cleared resource capacity.\par
 \begin{align}
 	\min CM = \sum_{r \in \mathcal{R}} C^{I}_{r} p^{CM}_{r} + \sum_{j \in \mathcal{J}} C^{I}_{j} p^{CM}_{j}
 	\label{LL_lower_CM_obj} 	
 \end{align} 
 subject to:
\begin{IEEEeqnarray}{l'l'l}
    \sum_{j \in \mathcal{J}} {D}^{th}_j = \sum_{r \in \mathcal{R}} p^{CM}_{r} + \sum_{j \in \mathcal{J}} p^{CM}_{j} & & [\lambda^{CM}]  \label{CM_eq} \\
    0 \leq p^{CM}_{r}  \leq \overline{{P}_{r}} {A}^{CM}_{r} u_r & \ \forall r \in \mathcal{R} & \label{CM_r} \\
    0 \leq p^{CM}_{j}  \leq {D}^{th}_j,\ & \forall j \in \mathcal{J} & \label{CM_j}
\end{IEEEeqnarray} \par
The objective function (\ref{LL_lower_CM_obj}) minimises costs. The first term represents total capacity costs as the product of investment costs $C^{I}_{r}$ and cleared resource capacity $p^{CM}_{r}$. The second term represents the costs of unmet capacity as the product of penalty costs associated with each capacity demand segment $C^{I}_{j}$ and the capacity shortage in each segment $p^{CM}_{j}$, where the dual ($\lambda^{CM}$) defines the marginal clearing price of the capacity auction.
Equation (\ref{CM_eq}) balances auction demand and supply. Equation (\ref{CM_r}) ensures that cleared capacity is limited by the derated maximum capacity of resource. Capacity deratings are based on the effective load carrying capacity (ELCC) \cite{schlag2020capacity}. Capacity demand curve segments are specified in (\ref{CM_j}) \cite{Aderhold2011}.\par  
\subsubsection{Investment decision}
A single representative agent is used to model investment decision making for each individual generation, hydro or storage resource. Each resource is modelled as a lumpy binary investment with risk endogenised via a risk-weighted utility function measured as a convex combination of expected value and a coherent risk measure in the decision problem for each resource \cite{Mays2019,Hanspeter2018}.\par
\begin{multline}
	\max U^G_r = \beta_r ({V}^G_r - \frac{1}{\alpha_r^G}\sum_{\omega\in\Omega} \pi_\omega \varrho^G_{g\omega} ) \\ + (1-\beta_r) \sum_{\omega \in \Omega} \pi_{\omega} {\Psi^G_{r\omega}} - C^{I}_{r} \overline{{P}_{r}} u_r \label{U_r1}
\end{multline}\par
subject to:
\begin{align}
	{\Psi^G_{r\omega}} = (\mathbf{\lambda}^E_{\omega n(r)} - \mathbf{C}^{vc}_r).\mathbf{p}^{G}_{r\omega} + ( \mathbf{\lambda}^{R}_{\omega}- \mathbf{C}^{R}_r). \mathbf{p}^{R}_{r\omega} + \lambda^{CM} p^{CM}_{g} \label{profs} 
\end{align}
\begin{align}
    {V}^G_r - \Psi^G_{r\omega} \leq \varrho^G_{r\omega},	\ \forall \omega \in \Omega \label{cvarcon_1} \\
	\varrho^G_{r\omega} \geq 0,	\ \forall \omega \in \Omega \label{cvarcon_2}     
\end{align}
The objective function (\ref{U_r1}) is specified as a maximization of risk-weighted utility, formulated as convex combination of the expected value and conditional value-at-risk (CVAR) of scenario profits (\ref{profs}), minus capital costs. Constraints (\ref{cvarcon_1}) and (\ref{cvarcon_2}) are required for the scenario formulation of CVAR \cite{Rockafellar2002}.\par 
\subsection{Insurance overlay} \label{:sec ins_methods}
This section sets out the decision making model for the insurer (${INS}$), and for consumers (${CON_d}$) given an insurance scheme. We assume the insurer is a central agent with contingent liability for consumer electricity service outages. We also assume the insurance is mandatory. These two assumptions are made primarily for technical convenience. While decentralized and competitive paradigms for insurance are possible, this magnifies computational complexity, which we will consider in future work.\par
\begin{equation}\label{ins_obj}
	\max  U^i = (1-\beta_i)\sum_{\omega \in \Omega} \pi_{\omega} \Psi^i_{\omega} + \beta_i \tilde{\mathit{c}}^i - \gamma \phi^i
\end{equation}
subject to:
\begin{multline}\label{profit_in}
	\Psi^i_{\omega} = \sum_{d \in \mathcal{D}}(C^P_d -  \mathbf{C^{comp}_d}.\mathbf{p^{c}_{d\omega}}) - \sum_{r\in\mathcal{R}^{der}} \kappa C^{I}_r \overline{P_{r}}, \ \forall \omega \in \Omega 
\end{multline}
\begin{IEEEeqnarray}{l}
	\tilde{\mathit{c}}^i ={V}^i - \frac{1}{\alpha^i}\sum_{\omega\in\Omega} \pi_\omega \varrho^i_{\omega} \label{cvarconin} \\ 
	{V}^i - \Psi^i_{\omega} \leq \varrho^i_{\omega},	\ \forall \omega \in \Omega \label{cvarconin3} \\
	\tilde{\mathit{c}}^i \geq -\phi^i \label{cvarin2} \\
	\overline{P_r} \geq 0, \varrho^i_{\omega}\geq 0,\phi^i\geq0,\mathbf{p^{c}_{d\omega}}\geq0,\mathbf{p^{d}_{d\omega}}\geq0 \label{IEEEeqnarray}
\end{IEEEeqnarray}
\begin{IEEEeqnarray}{l'l}
    \sum_{d\in\mathcal{D}^n} \mathbf{p}^{c}_{d\omega} = \sum_{d\in\mathcal{D}^n} \mathbf{p}^{sh*}_{d\omega} - \sum_{r\in\mathcal{R}^{der}} \mathbf{p}^{G}_{r\omega} \> \> \> \forall \omega \in \Omega n \in \mathcal{N} \label{p_curt} & \\
    0 \leq \mathbf{p}^{G}_{r\omega} \leq \overline{P_{r}} \mathbf{A}^G_{r\omega} \> \> \> \forall r \in \mathcal{R}^{der}, \omega \in \Omega \label{gen_ins} & \\
    0 \leq \mathbf{S_{r\omega}} \leq {\overline{{P}_{r}}}{e_{r}}  \> \> \> \forall r \in \mathcal{S}^{der},\omega \in \Omega & \label{soc_lim}
\end{IEEEeqnarray}
\begin{align}
	S_{rt\omega} = S_{r t-1 \omega} + q^{+}_{r} p^{G+}_{rt\omega} - \frac{1}{q^{-}_{r}} p^{G-}_{rt\omega} \forall r \in \mathcal{S^{der}} ,t \in \mathcal{T},\omega \in \Omega  \label{soc_def}
\end{align}
The objective function (\ref{ins_obj}) is a maximisation of the convex combination of the mean and CVAR ($\tilde{\mathit{c}}^i$) of the insurer's profits (the first and second term). In addition, the insurer must also bear the costs associated with reserving capital to meet potential losses \cite{Billimoria2022}. This is represented in the third term of the objective function as a product of the reserved capital ($\phi^i$) and the annualized cost of capital ($\gamma$). Insurer profits ($\Psi^i_{\omega}$) are defined in (\ref{profit_in}) as the sum of premium revenues, minus insurance compensation costs and the investment costs of RDER, scaled by the subsidy $\kappa$ provided to consumers.\par
CVAR is defined in (\ref{cvarconin}) with the auxillary CVAR constraint in (\ref{cvarconin3}). Equation (\ref{cvarin2}) sets out the reserving requirements for reserve capital to be in excess of the negative CVAR. Equation (\ref{p_curt}) defines load shedding as the difference between the wholesale unserved demand (an output of $ED_\omega$) minus generation from RDER. Technical constraints associated with RDER (availability, SoC etc) are set out in (\ref{gen_ins})-(\ref{soc_def}).\par
Finally we set out the decision making framework for consumers ($CON_d$) to allow the consumer to invest in  RDER at a subsidised cost.\par
\begin{equation}\label{cons}
	\max  U^c_d = (1-\beta_d)\sum_{\omega \in \Omega} \pi_{\omega} \Psi^c_{d\omega} + \beta_d \tilde{\mathit{c}}^c_{d}
\end{equation}
subject to:
\begin{multline}
	\Psi^c_{d\omega} = -\mathbf{C^{voll}_d} \mathbf{p}^{c}_{d\omega} - \sum_{r\in\mathcal{R}^{der}} (1-\kappa) C^{I}_r \overline{P_{r}} - C^P_d \\+ \mathbf{C}^{comp}_d \mathbf{p}^{c}_{d\omega}, \ \forall \omega \in \Omega \label{conscfs}
\end{multline}
\begin{IEEEeqnarray}{l'l}
	\tilde{\mathit{c}}_c^d = {V}_c^d - \frac{1}{\alpha_c^d}\sum_{\omega\in\Omega} \pi_\omega \varrho^c_{d,\omega} \label{consumercvar} \\
 {V}_c^d - \Psi^c_{d,\omega} \leq \varrho^c_{d,\omega},\ & \forall \omega \in \Omega \label{cvarconsumer} \\
	\varrho^c_{d,\omega} \geq 0	& \ \forall \omega \in \Omega \label{cvarconsumer2} \\ 
	\mathbf{p}^{c}_{d\omega} = \mathbf{p}^{sh*}_{d\omega} - \sum_{r\in\mathcal{R}^{der}} \mathbf{p}^{G}_{r\omega} & \forall \omega \in \Omega \label{p_curt_c}\\
	0 \leq \mathbf{p}^{G}_{r\omega} \leq \overline{P_{r}} \mathbf{A}^G_{r\omega} & \forall r \in \mathcal{R}^{der}, \omega \in \Omega \label{gen_cons}\\ 
 0 \leq \mathbf{S_{rt\omega}} \leq {\overline{{P}_{r}}}{e_{r}} & \forall r \in \mathcal{S}^{der},\omega \in \Omega \label{soc_lim_c}
\end{IEEEeqnarray}
\begin{multline}
	S_{rt\omega} = S_{r t-1 \omega} + q^{+}_{r} p^{G+}_{rt\omega} - \frac{1}{q^{-}_{r}} p^{G-}_{rt\omega} \\ \forall r \in \mathcal{S}^{der} ,t \in \mathcal{T},\omega \in \Omega  \label{soc_def_c}\\
\end{multline}
The objective function is a risk-weighted utility maximization of consumer surplus ($\Psi^c_{d\omega}$). The consumer surplus, as defined in (\ref{conscfs}), reflects losses associated with load shedding, less investment costs of RDER (net of subsidy), less the insurance premium plus any insurance compensation payable for load shedding. CVAR is defined in (\ref{consumercvar}). For each consumer, the key decision variable relates to the capacity of DER built ($ \overline{P_{r}}$). The remainder relate to CVAR (\ref{consumercvar})-(\ref{cvarconsumer2}) and technical/operational constraints (\ref{p_curt_c})-(\ref{soc_def_c}),similar to the ${INS}$ problem. 

\subsection{Market equilibrium algorithm} \label{:sec market_eq_alg}
We seek to find a market investment equilibrium where no agent can increase its utility by deviating unilaterally from the solution. To search for equilibria we propose a heuristic algorithm that seeks to replicate the process of competitive entry and exit in liberalised markets. Algorithm \ref{algo} shows the approach adopted.\par
\begin{algorithm}
	\SetAlgoLined
	\SetKwInOut{Input}{input}\SetKwInOut{Output}{output}
	\Input{Initial instance of resource capacities ($GMP_g$)}
	\Output{Equilibrium solution}
	initialization: set resource mix $u_r \forall r \in \mathcal{R}$,$\epsilon$, \\ iteration counts $x,y,z$;\ \\ 
	{
\textit{Market loop}:\\
		\While{$\max_{r \in \mathcal{R}}|u_{r,(x)} -u_{r,(x-1)} |>\epsilon$}{
			\textit{Retirement loop}:\\
			\While{$\max_{r \in \mathcal{R}^{ret}}|u_{r,(y)} -u_{r,(y-1)}|>\epsilon$}{
			{solve (${ED}_\omega \forall \omega \in \Omega$)}	\\
			{solve ($CM$)}	\\
			 \For{$r \in \mathcal{R}^{ret}$}{solve ($U^G_r$)
				}
				$u_{r,(y)} \leftarrow 0$ for $r \in \arg \min (U^G_r \forall r \in \mathcal{R}^{ret})$			
			}	
\textit{Investment loop}:\\
\While{$\max_{r \in \mathcal{R}^{inv}}|u_{r,(z)} -u_{r,(z-1)} |>\epsilon$}{
	\For{$r \in \mathcal{R}^{inv}$, {in interconnection queue position order}}{
		$u_{r,(z)} \leftarrow 1 $ 
		{solve (${ED}_\omega \forall \omega \in \Omega$)}	\\
		{solve ($CM$)}	\\
		solve ($U^G_r$) \\
		\uIf{$U^G_r > 0$}{
			$u_{r,(z)} \leftarrow 1 $ \;
		}
		\Else{
			$u_{r,(z)} \leftarrow 0 $ \;
		}
	}		
}
		}
}
solve (${INS}$)\\
solve (${CON_d} \forall d \in \mathcal{D}$) \\
	return
	\caption{Algorithm to find a market equilibrium}
	\label{algo}
\end{algorithm}
The algorithm commences with the current resource capacity mix, finding the implied dispatch solutions and prices for energy, reserves and capacity. Resource capacity updates take place in two loops, a \lq retirement' loop and a \lq new investment' loop, which are run sequentially and iterated. Given the potential for multiple equilibria especially given lumpy investment (binary variables), our algorithm is best described as a guided search through the feasibility set. The rationale of this approach is to seek an equilibrium that is intuitive in practice by the nature of retiring in order of un-profitability and investing by order of the interconnection queue. The algorithm terminates when the resource mix does not change over the prior iteration (i.e. no plants seek to retire and no new plants seek to enter the market). \par
For each iteration of the retirement loop, units in the current mix with a risk-weighted utility, $U^G_r < 0$ are considered for retirement. The unit with the lowest risk-weighted utility is retired and the loop re-iterates and updates prices and dispatch. The process continues until there are no more units with a negative risk-weighted utility. \par
In the investment loop, new candidate resources are considered for investment by the order of their position in the interconnection queue.  Where the interconnection queue is not transparent reasonable estimates could be made through publicly available connections information, such as the categorisation of commitment status (see \cite{AEMO2022}). A resource is selected for investment if it has a non-negative risk-weighted utility when added to the resource mix (with updated prices and dispatch). The loop continually iterates, terminating when there are no new candidate units added to the mix. This process ensures that each new investment made is viable (i.e. has a non-negative risk-weighted utility) on a stand alone basis. Once a market equilibrium is reached under steps (1)-(26), we then apply the insurance and consumer decision-making model sequentially (as they do not interfere with market outcomes) with inputs for consumer outages based on the results of the market equilibrium. \par
\section{Numerical Study} \label{:sec numerical}
Our numerical study is developed to illustrate the insurance value of resilient investment. The National Electricity Market (NEM) of Australia provides an apposite case study of a large scale grid in transition towards a high penetration of VRE and a rolloff of legacy fossil fleet. We also benefit from a high degree of transparency on demand and generation availability projections across scenarios and locations, technical and financial data for current fleet and project interconnection pipeline and network topology information.
\subsection{Data and assumptions}
Plant technical, financial and cost data are sourced from the Integrated System Plan (ISP) produced by the Australian Energy Market Operator (AEMO) \cite{AEMO2022a} for existing, committed and anticipated resources, supplemented by  \cite{AEMO2022} for new projects. For the network topology we adopt the ISP sub-regional network representation comprising of 10 zones, with specific network transfer capability and seasonal availability limits  \cite{AEMO2022a}.\footnote{Supplementary Information Section II provides a diagram of the topology and flow limits.} \par
To account for weather uncertainty, a set of annual weather-year scenarios are adopted for demand, variable renewable generation availability, hydro inflows and transmission network capacity with traces provided for every half-hour over the year. Projections from ten equiprobable 'base' weather years reflect normal weather variability as sourced from AEMO's ISP Step Change projection. AEMO's weather year forecasts are developed from ensemble projections of downscaled global climate models and reflect inherent correlations between parameters such as demand and variable renewable generation availability. Twenty four representative days are selected from each of the base scenarios using a K-means clustering algorithm \cite{Jenkins2017}. We approximate energy exchange of long duration storage and hydro between representative periods through the introduction of additional variables and constraints based on the approach in \cite{Sepulveda2021}. In order to assess the impact of extreme outcomes, the base weather years are complemented with six equiprobable 'extreme' years, developed as stylised scenarios that reflect the specific risks facing the NEM building upon extreme scenario calibration work undertaken in \cite{AEMO2020d} and the Electricity Sector Climate Information Risk Assessment Framework \cite{AEMO/CSIRO2021} (specific assumptions used are provided in Supplementary Information Section III).  They serve to illustrate the range of extreme events that could be expected to form an insurance-based assessment of extreme risks in practice. A real-world analysis would involve a larger number of scenario assessments, which we have limited here for computational tractability. Each of the six extreme year scenarios is assumed to have a probability of occurrence of 0.01 (i.e. each a 1-in-100 year event). Appendix A provides further details on the development of the extreme scenarios. \par
Three market designs are tested in the case study - (i) an energy only market (EOM) design, (ii) an energy market with an operating reserve demand curve (ORDC) and (iii) an energy market with capacity auctions (CM). An energy market price cap of \$15000/MWh is adopted for the EOM and ORDC designs, while for the CM we adopt a reduced energy market price cap of \$2000/MWh. A system-wide ORDC with three reserve quantity segments of 2000 MW, 1000 MW and 1000 MW with corresponding price thresholds of \$15000/MWh, \$10000/MWh and \$5000/MWh. The CM adopts a capacity demand curve with three interpolated points corresponding to specified proportions of maximum demand of 0.95, 0.05 and 0.05. Corresponding capacity price thresholds are based on an assumed cost of new entry (CONE) of \$90000/MW/year and set at 0.5, 1.0 and 1.5 times CONE respectively. The derating factors for numerical study are based on a marginal effective load-carrying capacity (ELCC) methodology.\par
Resources are assumed to have $\beta_r$ of 0.5 and tail probability $\alpha_r$ of 0.90. The insurance scheme adopts a capital reserving threshold with a tail probability $\alpha_i$ set at 99\% (consistent with international insurer solvency standards \cite{Eling2007}). In the base case we assume that the insurance utility preferences are skewed towards expected returns i.e. a $\beta_i$ of 0.1, which provides a conservative estimate of the potential benefits of the insurance scheme. \par
%In the base case retail consumers are assumed to be risk neutral with $\beta_d$ of 0 and CVAR confidence level $\alpha_d$ of 0.99. This is consistent with the notion that consumers may systematically underestimate extreme risks. Once again this is intended to provide a conservative estimate of scheme value, which is also sensitised against higher levels of consumer risk aversion. 
We provide a set of resilient distributed energy resource (RDER) investment options for the insurer. The insurer is able to select from a combination of resources that comprise rooftop solar and distributed battery storage (costs and technical specifications from \cite{prasanna2021storage}). In this exercise we consider both (i) a direct investment model where the insurer directly funds the investment and bears associated costs and(ii) a subsidy model, where capital subsidies are provided to consumers for the deployment of RDER storage. We focus on storage subsidies only, given the array of subsidies available to distributed solar technologies.
\subsection{Results}
\begin{figure*}%[t]
 	\centering
 	\includegraphics[height=0.65\columnwidth]{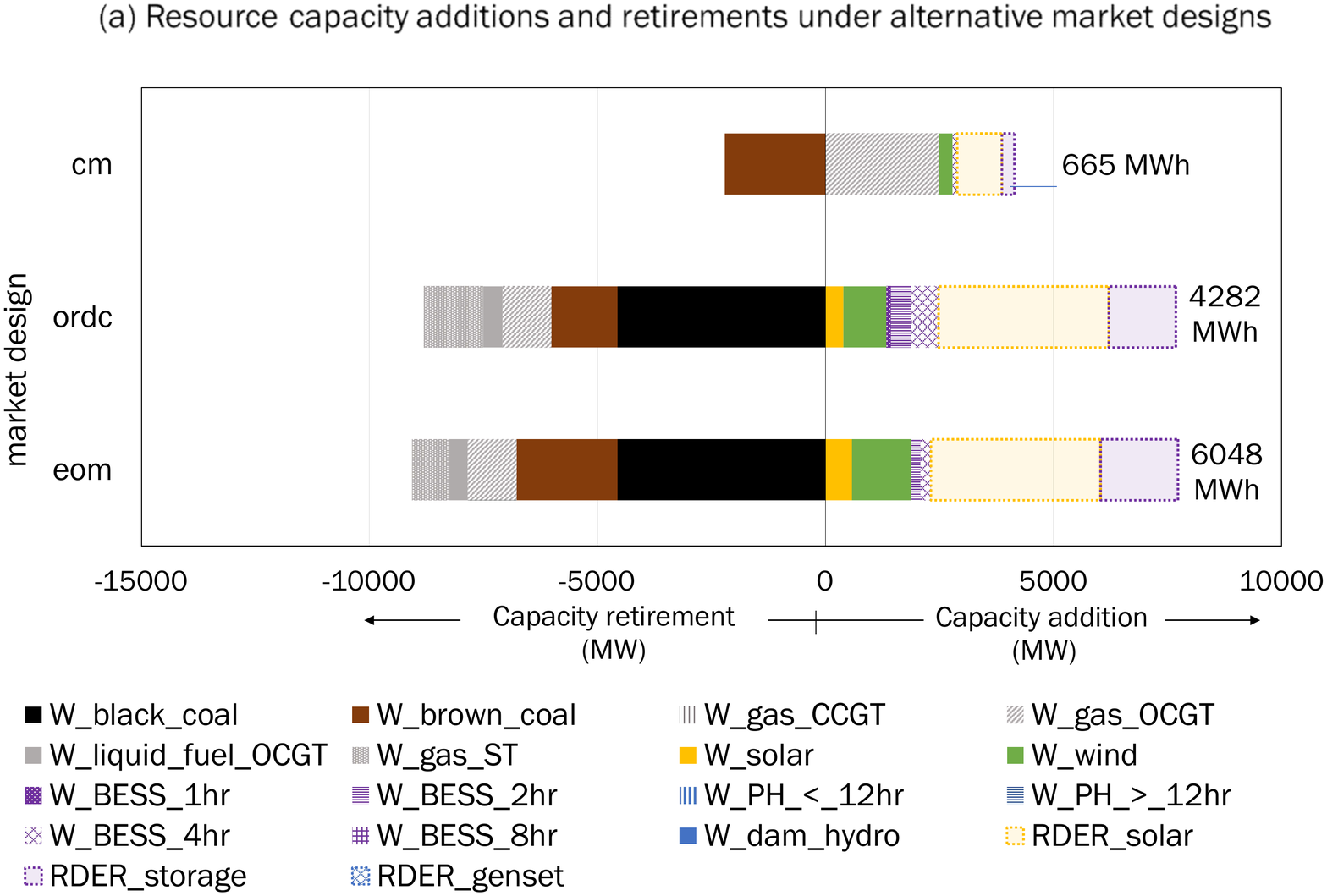} 
 	\includegraphics[height=0.65\columnwidth]{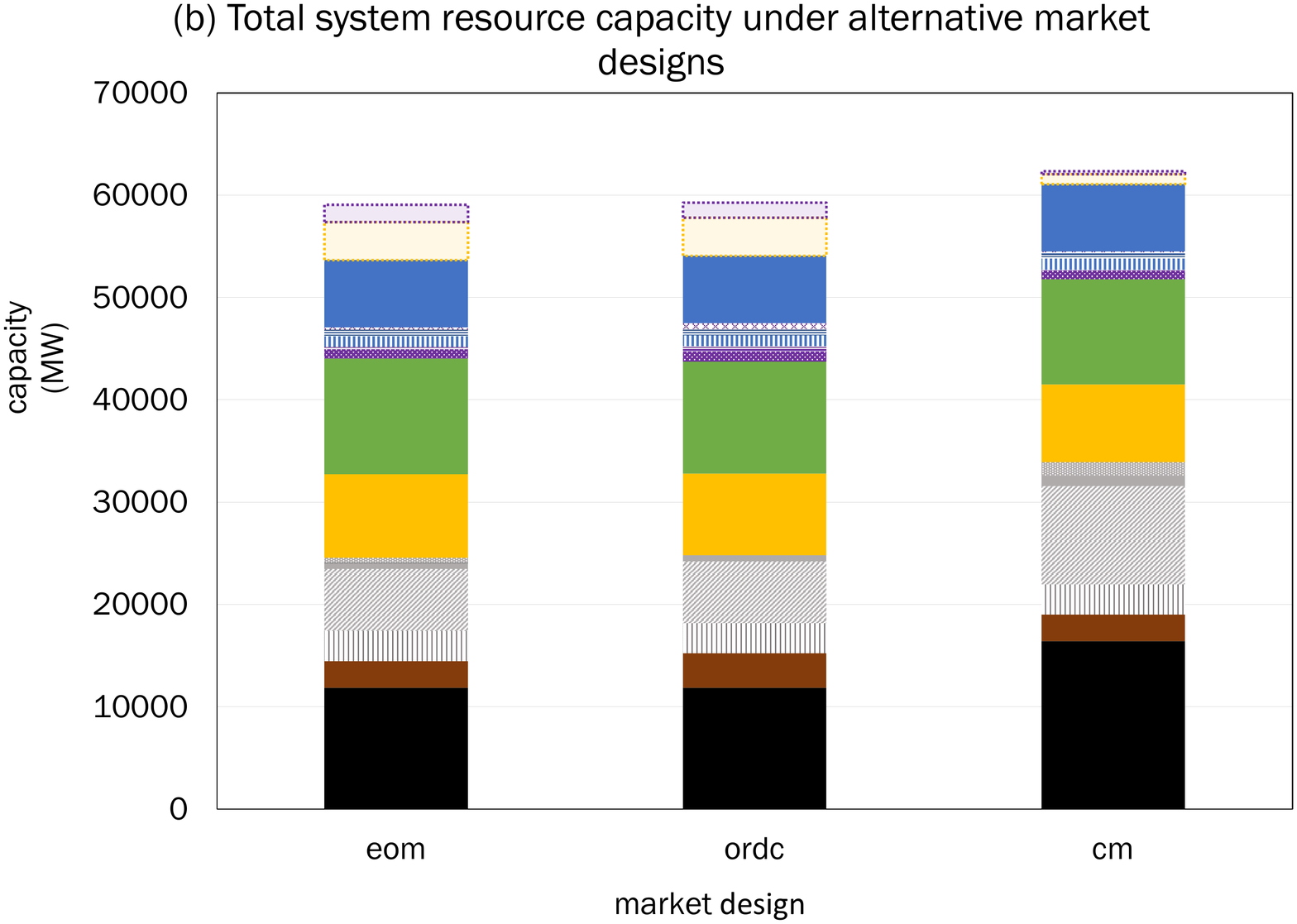} 
 	\caption{Resource capacity additions and retirements, panel (a), and total resource capacity, panel (b), under alternative market designs. Storage durations in MWh are also indicated panel (a) }
 	\label{fig:capacity_MW}
\end{figure*}

\begin{figure}%[t]
	\centering
	\includegraphics[height=0.70\columnwidth]{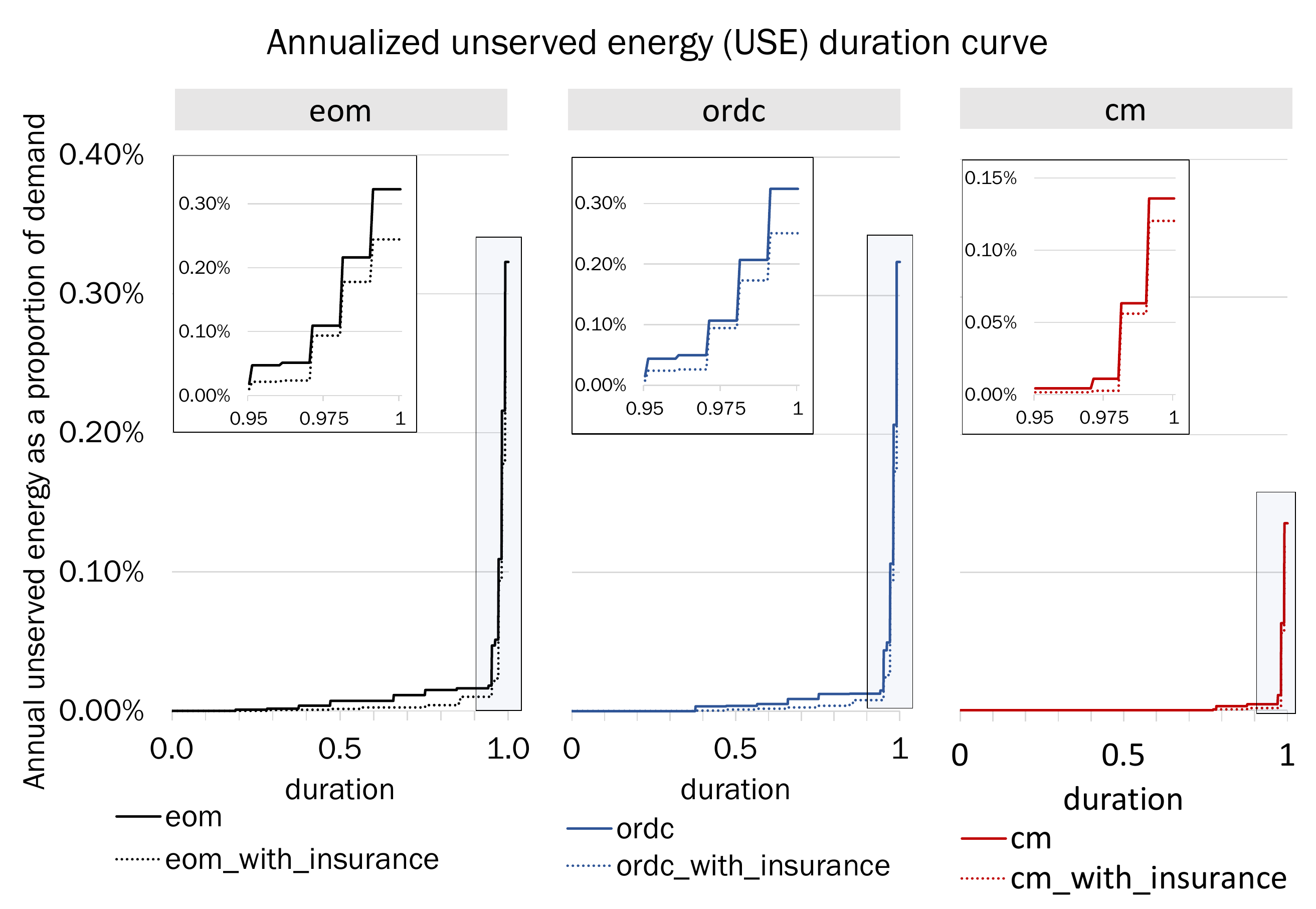} 
	\caption{Duration curves for system unserved energy (USE), measured as percentage of demand, under alternative market designs}
	\label{fig:tot_use}
\end{figure}

Figure \ref{fig:capacity_MW} panel (a) illustrates retired and added capacity, and panel (b) shows total installed capacity at the system level under each of the three selected market designs. The figures illustrate both the capacity incentivised by the wholesale market design (resource categories preceded with ``W''), as well as additional investment in resilient DER resource capacity funded by the insurance scheme (preceded by ``RDER''). Duration curves for system annualized unserved energy (USE) as a proportion of demand are shown in Figure \ref{fig:tot_use} for each of the three market designs.
The results indicate differences in both the total quantities and type of resources incentivised by each of the market designs. At a wholesale level and relative to current supply mix, the CM design results in a net addition of 0.7GW of resource capacity, while the EOM and ORDC designs drive net retirements of over 6.7GW and 6.3GW respectively. In the spot based designs (EOM and ORDC) retirements (of ~9.0GW) are mainly from black \& brown coal, but also some gas units with new investment in the form of wind, solar and storage (of 1 and 2 hour durations). New investment in the CM comes from fast-start gas units and some storage (though the latter is incremental, as all candidate gas units in the current queue are built). At a wholesale level Figure \ref{fig:tot_use} illustrates that the base reliability outcomes (prior to the application of the insurance scheme) are better for CM relative to an EOM and ORDC across median and high exceedance probabilities. This is expected given the modelling of the designs and the objectives of a CM that is targetted towards maximal load forecasts.\par
\begin{figure*}%[t]
	\centering
	\includegraphics[height=0.5\columnwidth]{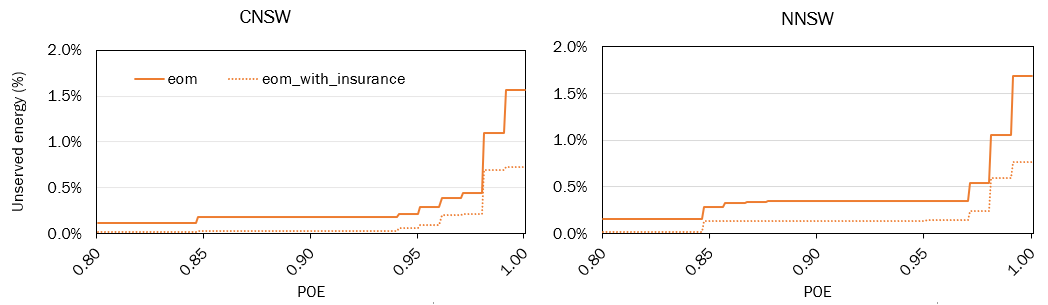} 
	\caption{Duration curves for unserved energy (USE), measured as percentage of demand, in regional areas of Central New South Wales (CNSW) and Northern New South Wales (NNSW) under an energy only market design}
	\label{fig:eom_use_regional}
\end{figure*}
\begin{figure}%[t]
	\centering
	\includegraphics[height=0.80\columnwidth]{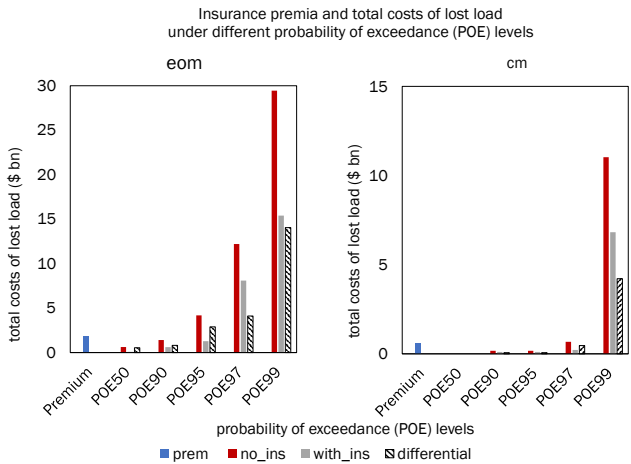} 
	\caption{A comparison of the insurance premia required for viability of the insurance scheme relative to the reduced cost burden of lost load borne by consumers}
	\label{fig:costs}
\end{figure}

The impact of the insurance framework on resilience is evident in the quantity of RDER that the insurance agency is incentivised to build, and consequently unserved energy outcomes. For the EOM and ORDC, the insurance scheme drives additional investments of 3.7 GW in RDER-solar and ~1.5-1.7GW in RDER-storage (with an average duration of 3-4 hours). For the CM ~1GW of RDER-solar and 0.3GW of RDER-storage (2 hr duration) are incentivised. In terms of reliability, improvements in unserved energy outcomes for extreme cases are observed for all market designs as a result of the additional investment in resilient DER. At a probability of exceedance (POE) level of 95\%, USE is improved by ~0.007-0.008\% for EOM/ORDC and 0.003\% for CM, while for POE of 99\%, improvements recorded are 0.034-0.038\% for EOM/ORDC and 0.007\% for CM above the wholesale market outcomes. Local effects are also prominent, with regional areas with weaker connections to the network experiencing poorer reliability outcomes, such as Central New South Wales (CNSW) and Northern New South Wales (NNSW). The potential contingent liability exposures investment under an insurance framework is skewed to such regions. Particularly in the EOM case investments in RDER drives improvements in USE outcomes in regional areas (see Fig. \ref{fig:eom_use_regional}).
In terms of comparing the relative efficiency of the scheme, Figure \ref{fig:costs} illustrates the total costs of lost load (effectively the product of the unserved energy in MWh terms and the value of lost load) under various POE levels for the case with no insurance and the case with insurance. The differential between the two cases indicates the reduction in outage costs that can be achieved with the modelled insurance scheme.  This is compared against the total insurance premium required to be levied to maintain a solvent insurer (i.e. with a zero risk measure). For example, for the EOM case, given a required premium of \$1.8 billion, outage costs are reduced by \$0.8bn for a POE90 level, increasing to \$2.9 billion for POE95 and \$14.0 billion for POE99. For the CM, the insurance scheme becomes cost-advantageous at levels between POE97 and POE99.\par
\begin{figure}[t]
	\centering
	\includegraphics[height=0.70\columnwidth]{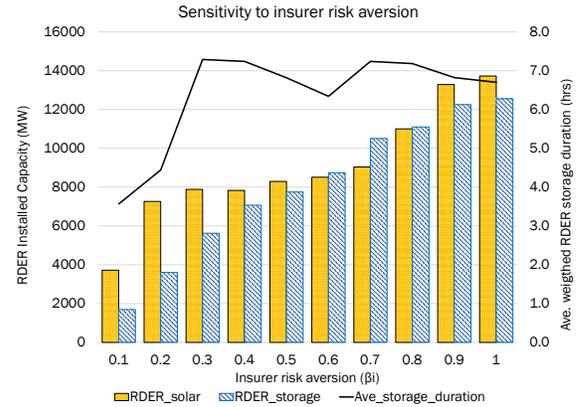} 
	\caption{Sensitivity of installed RDER generation and storage capacity to risk aversion of insurer}
	\label{fig:eom_risk_sens}
\end{figure}

We also undertake sensitivities around the level of insurer risk aversion with results for the EOM design shown in Figure \ref{fig:eom_risk_sens}. Our base case is conducted on the assumption of low insurer risk aversion ($\beta_i=0$), though as assumed risk aversion increases we observe that the level of RDER (both solar and storage) investment grows significantly with risk aversion (with roughly double the quantity of investment at $\beta_i=0.5$ and over 3 times under a fully risk-averse case , $\beta_i=1.0$. The average weighted duration of storage also tends to increases with risk aversion from 3-4 hours and to 6-7 hours (fluctuating at those levels for $\beta_i\geq 0.3$).\par
\begin{figure}%[t]
	\centering
	\includegraphics[height=0.7\columnwidth]{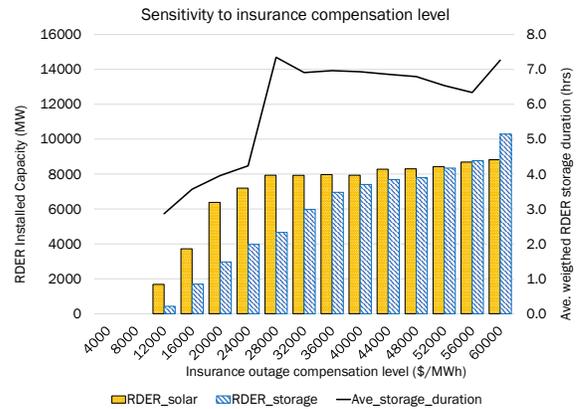} 
	\caption{Sensitivity of installed RDER generation and storage capacity to insurance outage compensation levels}
	\label{fig:eom_comp_sens}
\end{figure}

Further a sensitivity analysis is also conducted against insurance compensation levels with results for the EOM design shown in Figure \ref{fig:eom_comp_sens}. The results indicate that the insurance scheme fails to incentivise investment in RDER at compensation levels below \$12000/MWh. Beyond this  level, RDER investment grows but starts to cap out at maximal compensation levels of ~\$28000/MWh. This indicates that there are practical bounds to the value of the insurance scheme in a large scale market context.\par
\begin{figure}%[t]
	\centering
	\includegraphics[height=0.73\columnwidth]{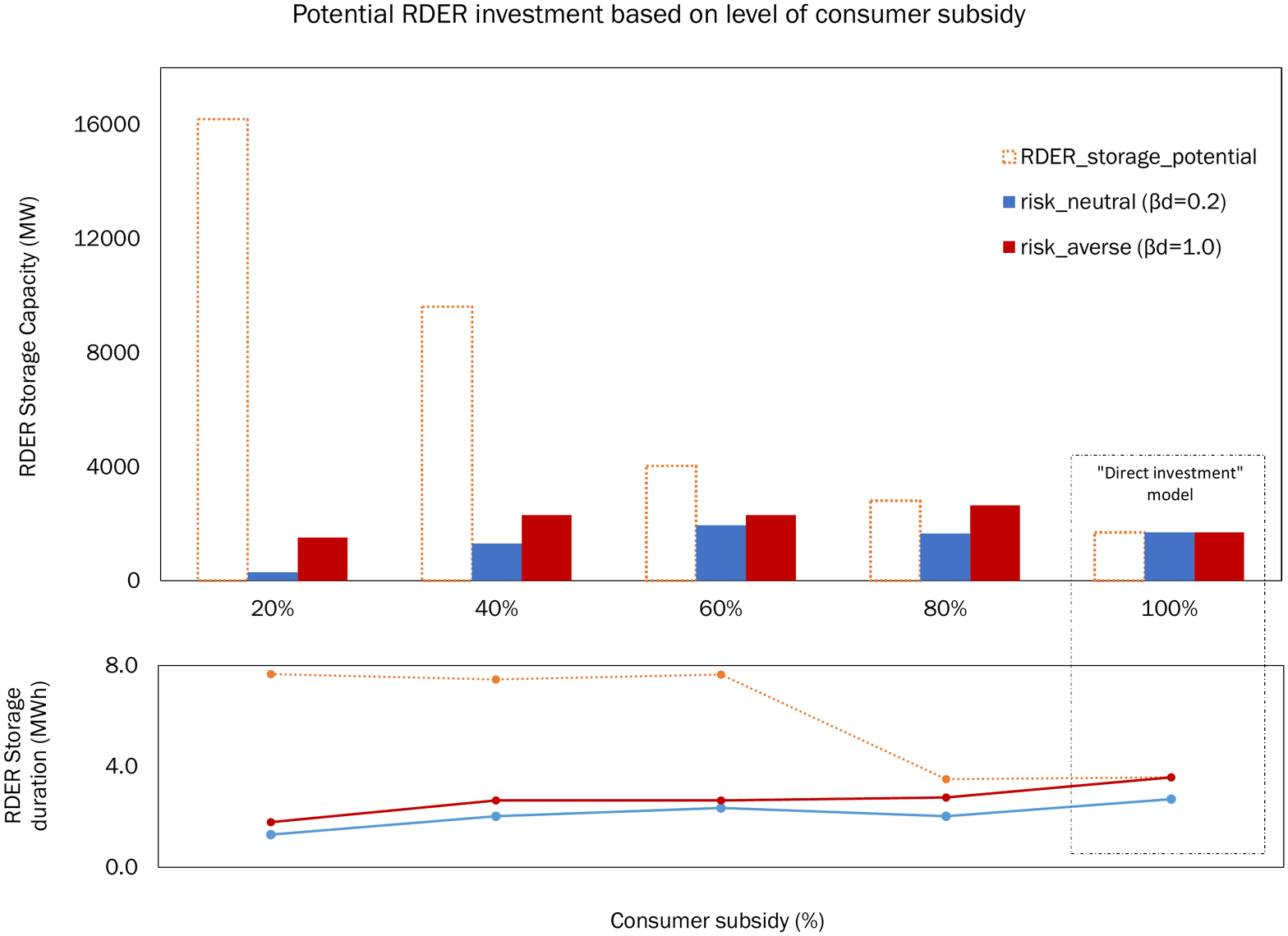} 
	\caption{Consumer subsidy model: sensitivity of installed RDER storage capacity to subsidy levels}
	\label{fig:eom_subsidy_sens}
\end{figure}

Finally, while the above results are obtained under the assumption of direct investment by the insurer in RDER, we also consider a case where the insurer provides a subsidy to consumers ranging from 20\% to 100\% of the capital costs of RDER-storage under an EOM design. Figure \ref{fig:eom_subsidy_sens} illustrates two sets of data points - the first being the potential quantities of RDER storage investment viable from the perspective of the insurer. This effectively serves as a cap on the level of investment. As would be expected, the level of potential investment declines with higher subsidy levels. The second data set represents actual investments by consumers given the level of subsidy provided by the insurance scheme and the risk preferences of the consumer. We show results for consumers with near risk-neutral preferences (with $\beta_d=0.2$, and consumers that are risk averse (with $\beta_d=1.0$). Two points are evident. First, while the potential for investment is significant at lower subsidies (mitigating insurer investment costs) these levels are never realised given consumer budgets. Second, while subsidy levels of 60-80\% can drive higher investment relative to the 'direct investment' model, these levels are only realised with higher levels of consumer risk aversion. Finally, there are also significant differences between the storage duration appetite between the insurer and consumers at lower subsidy levels, though this gap narrows somewhat as subsidies increase. \par
\section{Policy Implications and Conclusions} \label{sec: conc}
There are a number of important policy implications and further areas of inquiry flowing from the results of the case study. First, the wholesale market outcomes reinforce the notion that extreme events present real risks for a power and energy system, with particular effects on consumers in poorly connected, remote regions of the grid. This effect remains evident in market designs that incentivize higher levels of investment, such as designs with capacity mechanisms. It is also interesting that outcomes of modelling the latter in a large scale grid remains consistent with the inherent bias of such market designs towards low capital/high marginal cost resources (such as legacy thermal generation), which can be detrimental in scenarios where thermal failure represents the extreme risk \cite{Mays2019,Mays2021b}. The need for resilience in large scale power systems thus remains an important objective. \par
Second, the insurance framework provides an economic lens for investment decision making particularly as it relates to high-impact lower probability events. Importantly the investment procurement quantities and required premia adjust to the capacity mix yielded by the market design. The results support a rational economic case for investment in resilience by central agencies.  Relatedly the cost effectiveness of the insurance scheme can be observed in its mitigation of consumer welfare losses in extremes (as well as median cases in the EOM) relative to the required premium. While the results do not suggest that all adverse outcomes can be avoided, it provides material economic protection to consumers, through the combination of economic loss compensation and loss-mitigating investment. This represents an important alignment with the policy objectives of system resilience which calls for improved resistance and adaptability rather than elimination of extreme impacts altogether \cite{NationalAcademiesofSciences2017}.\par
In terms of the scheme structure and organisational design, we observe that the risk parameters can have a material impact on the level of investment - and given the public nature of the insurance scheme, this would be an important area of consultation and engagement prior to implementation. Furthermore, the results also reflect the tradeoffs between insurer \lq subsidy' and \lq direct' modes of investment. Subsidisation models offer the potential for scalable investment, but dependent upon consumer risk attitudes and takeup (which may be time variant and subject to consumer budget constraints). The insurer has more control over direct investments but must bear all the costs, resulting in lower investment. Granular assessment of consumer attitudes and budgets should accompany any implementation. Finally the sensitivities also suggest that there are thresholds to scheme operation. With investment benefits only apparent within certain ranges, agencies would need to consider whether they are willing to meet minimum compensation levels over a long term basis.  The success of an insurance scheme depends upon its sustainability both from a capital and income perspective and as such should be considered as part of a programmatic approach to system resilience.\par
Going forward we consider that this paper supports further development of the research thesis. The funding of such a scheme requires attention to the economic willingness to pay and social acceptance of premiums to protect and compensate for losses, which are currently all borne by the consumer \cite{Baik2020}. The consideration of equity issues related to the allocation of such premiums is an important methodological stream, given that vulnerable consumers can often be located in the regions where risk is highest. The literature on equitable charging of tariffs is a natural starting point here \cite{borenstein2021designing}. Furthermore, government contingent liability is currently an open area of exposure. Comprehensive risk management standards relating to such exposures could aid in developing mitigation and investment plans for resilience. Finally, related streams could look to scheme design and optionality and whether micro-models of insurance could be applied at community levels.\par
The need for resilience in electricity systems is apparent and immediate. While wholesale market designs should be optimised for resilience, improvements to resilience can also come from distributed architectures, especially in the continuity of essential services during extreme weather. In our proposal for a social insurance scheme for electric service interruptions we align incentives for capital investment, and provide consumers with physical and financial risk mitigation. We illustrate that this can have material positive impacts in terms of encouraging DER investment, reduction of unserved energy during extremes, while providing financial coverage for consumers.\par
\bibliographystyle{IEEEtran}
\bibliography{IEEEabrv,library}
\end{document}

\bstctlcite{IEEEexample:BSTcontrol}
\title{Supplementary Information 
}

\author{\IEEEauthorblockN{Farhad Billimoria, Filiberto Fele, Iacoppo Savelli, Thomas Morstyn and Malcolm McCulloch}
\IEEEauthorblockA{\IEEEauthorrefmark{1}Department of Engineering Science, University of Oxford}
\IEEEauthorblockA{\IEEEauthorrefmark{2}School of Engineering, University of Edinburgh}\IEEEauthorblockA{\IEEEauthorrefmark{3}Centre for Research on Geography, Resources, Environment, Energy \& Networks, Bocconi University}
}

\maketitle
\section{Nomenclature} \label{sec:nomen}
This section sets out relevant nomenclature for the mathematical formulation:\par

\begin{table} [H]%[thpb]
	\caption{Model indices and sets}
	\renewcommand\arraystretch{1.3}
	%\begin{center}
		\begin{tabular}{p{0.2\linewidth}  p{0.7\linewidth}}%{ll}
			\hline
			Notation & Description \\
			\hline
			$r \in \mathcal{R}$ & $r$ denotes a resource and $\mathcal{R}$ is the set of all resources \\     
$\mathcal{G} \subset \mathcal{R}$	& $\mathcal{G}$ is the subset of all generation resources \\
$\mathcal{S} \subset \mathcal{R}$	& $\mathcal{S}$ is the subset of all storage resources \\
$\mathcal{H} \subset \mathcal{R}$	& $\mathcal{H}$ is the subset of all hydro generation resources \\
$\mathcal{G}^n,\mathcal{S}^n,\mathcal{H}^n \subset \mathcal{R} $ & The subset of all  resources, generation and hydro located at node/region $n$\\
$\mathcal{R}^{der}\subset \mathcal{R}$ & $\mathcal{R}^{der}$ is the subset of all resilient distributed energy resources available for investment by the insurer\\
$\mathcal{G}^{der}\subset \mathcal{R}$ & $\mathcal{G}^{der}$ is the subset of all resilient distributed generation resources available for investment by the insurer\\
$\mathcal{S}^{der}\subset \mathcal{R}$ & $\mathcal{S}^{der}$ is the subset of all resilient distributed storage resources available for investment by the insurer\\
$m,n \in \mathcal{N}$ & $n$ denotes a zone/node and $\mathcal{N}$ is the set of all zones/nodes in the network ($m$ is an alternate index)\\
   $l \in \mathcal{L}$	& $l$ denotes a transmission line and $\mathcal{L}$ is the subset of all transmission lines in the network \\
   $\mathcal{L}^{n}\subset \mathcal{L}$ & $\mathcal{L}^{n}$ is the subset of all transmission lines originating from node $n$\\
$d \in \mathcal{D}$ & $d$ denotes a consumer and $\mathcal{D}$ represents the set of all consumers\\
$\mathcal{D}^n \subset \mathcal{D} $ & The subset of all consumers located at node/region $n$\\
$i \in \mathcal{I}$ & $i$ denotes a segments used in the piecewise approximation of the operating reserve demand curve, and $\mathcal{I}$ is the set of segments\\
$j \in \mathcal{J}$ & $j$ denotes a segments used in the piecewise approximation of the capacity mechanism demand curve, and $\mathcal{J}$ is the set of segments\\
$\omega \in \Omega$ & $\omega$ denotes a scenario and $\Omega$ represents the set of scenarios\\
$t \in \mathcal{T}$ & $t$ denotes a dispatch interval (a half-hour) and $\mathcal{T}$ represents the set of all dispatch intervals\\
\hline
		\end{tabular}
	%\end{center}
	\label{tab:indices_sets}
\end{table}

\begin{table} [H]%[thpb]
	\caption{Parameters}
	\renewcommand\arraystretch{1.3}
	%\begin{center}
		\begin{tabular}{p{0.2\linewidth}  p{0.7\linewidth}}
			\hline
			Notation & Description \\
			\hline
$\alpha^G_r$ & $\alpha$-tail probability of the conditional-value-at-risk for market resources\\
$\alpha^i$ & $\alpha$-tail probability of the conditional-value-at-risk for the insurer\\
$\alpha^d_c$ & $\alpha$-tail probability of the conditional-value-at-risk for consumer $c$\\
$\beta_{r/d/i}$	& weight given to the CVAR for market resource $r$, consumer $d$, and insurer $i$\\
$\pi_{\omega}$	& The probability of scenario $\omega$\\
$C^{vc}_{rt\omega}   \>  [\mathbf{C^{vc}_{r \omega}}]$ & The short-run variable cost of energy delivered from resource $r$ at time $t$ in scenario $\omega$\\
$C^{I}_{r}$ & The annualised investment cost of resource $r$\\
$C^{R}_r$  & The short-run variable cost of providing reserve from resource $r$ at time $t$ in scenario $\omega$ \\
$C^{rsh}_i$ & The system penalty cost of unmet reserve for operating reserve demand curve segment $i$\\
$C^{sh}_d$	& The system value of lost-load for demand $d$ \\
$C^{I}_{j}$	& The administrative penalty cost of unmet capacity reserve for capacity mechanism demand curve segment $j$\\
$\overline{{P}^D_{dt\omega}} \>[\overline{\mathbf{P}}^D_{d\omega}]$ & Consumer energy demand at time $t$ in scenario $\omega$\\
$B_{nm}$ & The admittance of the transmission line from node $m$ to $n$.\\
${A}^G_{rt\omega} \> [\mathbf{{A}^G_{r\omega}}]$ & The generation availability of resource $r$ at time $t$ in scenario $\omega$\\
${A}^{CM}_{r}$ & The derated capacity of resource $r$ for the capacity mechanism auction, based on the effective load carrying capacity.\\
$A_{nm,t,\omega} \> [\mathbf{A^L_{nm,\omega}}]$ & Availability of the transmission line from node m to n \\
$\overline{P}_r$ & The power capacity of resource $r$\\
${u}_r$ & The binary build status of resource $r$, taking a value of 0 or 1\\
$q^{+}_r$ & The charging efficiency of storage resource $r$\\
$q^{-}_r$ & The discharging efficiency of storage resource $r$\\
$i^{G+}_{rt\omega}$ & Inflows to hydrological storage reservoir for resource $r$ at time $t$ in scenario $\omega$\\
$e_{r}$ & The maximum energy storage duration of resource $r$\\
$R^{req}_i$ & The required reserve in MW for operating reserve demand curve segment $i$ \\
$D^{th}_j$ & The required capacity demand in MW for capacity mechanism demand curve segment $j$ \\
$p^{sh*}_{dt\omega} \>[\mathbf{p^{sh*}_{d\omega}}]$ & Unserved energy of consumer $d$ at time $t$ for scenario $\omega$, as an output from the market equilibrium.\\
$\gamma$ & Annualised discount factor for capital investments\\
$\kappa$ & Capital investment cost subsidy offered to consumers by the insurer for RDER investments\\
$C^P_d$ & Insurance premium levied upon consumer $d$\\
$C^{voll}_d$ & The consumers value of reliability for load shedding for consumer $d$ \\
$C^{comp}_d$ & The insurance compensation payout value for consumer $d$ in \$ per MWh.\\
\hline
		\end{tabular}
	%\end{center}
	\label{tab:indices_sets}
\end{table}
* The vectorized form of the parameters that vary over time are shown in \textbf{bold} and square brackets.

\begin{table} [H]%[thpb]
	\caption{Decision variables}
	\renewcommand\arraystretch{1.3}
	%\begin{center}
		\begin{tabular}{p{0.2\linewidth}  p{0.7\linewidth}}
			\hline
			Notation & Description \\
			\hline
$p^{G}_{rt\omega} \> [\mathbf{p^{G}_{r\omega}}]$ & The energy dispatch of resource $r$ at time $t$ in scenario $\omega$.\\
$p^{G+}_{rt\omega} \> [\mathbf{p^{G+}_{r\omega}}]$& The energy discharge dispatch of storage resource $r$ at time $t$ in scenario $\omega$.\\
$p^{G-}_{rt\omega} \> [\mathbf{p^{G-}_{r\omega}}]$& The energy charge dispatch of storage resource $r$ at time $t$ in scenario $\omega$.\\
$p^{sh}_{dt\omega} \> [\mathbf{p^{sh}_{d\omega}}]$& The unserved load of consumer $d$ at time $t$ in scenario $\omega$.\\
$S_{rt\omega} \> [\mathbf{S_{r\omega}}]$ & The state of charge of storage or hydro resource $r$ at time $t$ in scenario $\omega$.\\
$p^{R}_{rt\omega} \> [\mathbf{p^{R}_{r\omega}}]$ & The reserve dispatch of resource $r$ at time $t$ in scenario $\omega$.\\
$p^{R+}_{rt\omega} \> [\mathbf{p^{R+}_{r\omega}}]$& The reserve discharge dispatch of storage resource $r$ at time $t$ in scenario $\omega$.\\
$p^{R-}_{rt\omega} \> [\mathbf{p^{G-}_{r\omega}}]$& The reserve charge dispatch of storage resource $r$ at time $t$ in scenario $\omega$.\\
$p^{rsh}_{it\omega} [\mathbf{p^{rsh}_{i\omega}}]$ & The unmet reserve for operating reserve demand curve segment $i$ at time $t$ for scenario $\omega$\\
$p^{CM}_{r}$ & The cleared capacity of resource $r$ for the capacity mechanism auction \\
$p^{CM}_{j}$ & The unmet quantity of capacity market demand for capacity mechanism demand curve segment $j$\\
$\theta_{t\omega n} [\mathbf{\theta_{\omega n}}]$ & The phase angle of node $n$ at time $t$ for scenario $\omega$ \\
$\lambda^E_{t\omega n} [\mathbf{\lambda^E_{\omega n}}]$	& Locational marginal price for energy for node $n$ at time $t$ for scenario $\omega$\\
$\lambda^R_{t\omega } [\mathbf{\lambda^R_{\omega }}]$ & The system marginal price for operating reserve at time $t$ for scenario $\omega$\\
$\lambda^{CM}$ & The system marginal price for capacity based on the clearing of the capacity mechanism.\\
$\overline{{P}_{g}}$ & The resource capacity for DER resource $g$ in the ${INS}$ and ${CON}_d$ problems. \\
${V}_G^r$ &	Auxiliary decision variable representing value-at-risk for market resource $r$\\
${V}^i$	& Auxiliary decision variable representing value-at-risk for insurer $i$\\
${V}_c^d$ & Auxiliary decision variable representing value-at-risk for consumer $d$\\
$\varrho^G_{g\omega}$ &	CVAR auxiliary decision variable as positive difference between $z_r^G$ and scenario profits for resource $r$ \\
$\varrho^i_{\omega}$ & CVAR auxiliary decision variable as positive difference between $z^i$ and scenario profits for insurer \\
$\varrho^c_{d\omega}$	& CVAR auxiliary decision variable as positive difference between $z_d^c$ and scenario profits for consumer $d$\\
$\phi^i$ & Required capital reserves for insurer\\
${p}^{c}_{d,t,\omega} [\mathbf{p}^{c}_{d\omega}]$ & Quantity of load shed  for consumer $d$ at time $t$ for scenario $\omega$ \\
\hline
		\end{tabular}
	%\end{center}
	\label{tab:indices_sets}
\end{table}
* The vectorized form of the decision variables that vary over time are shown in \textbf{bold} and square brackets.\par
\section{Modelled network topology}  \label{sec:App_A}
\begin{figure}[H]
	\centering
	\includegraphics[height=\textheight]{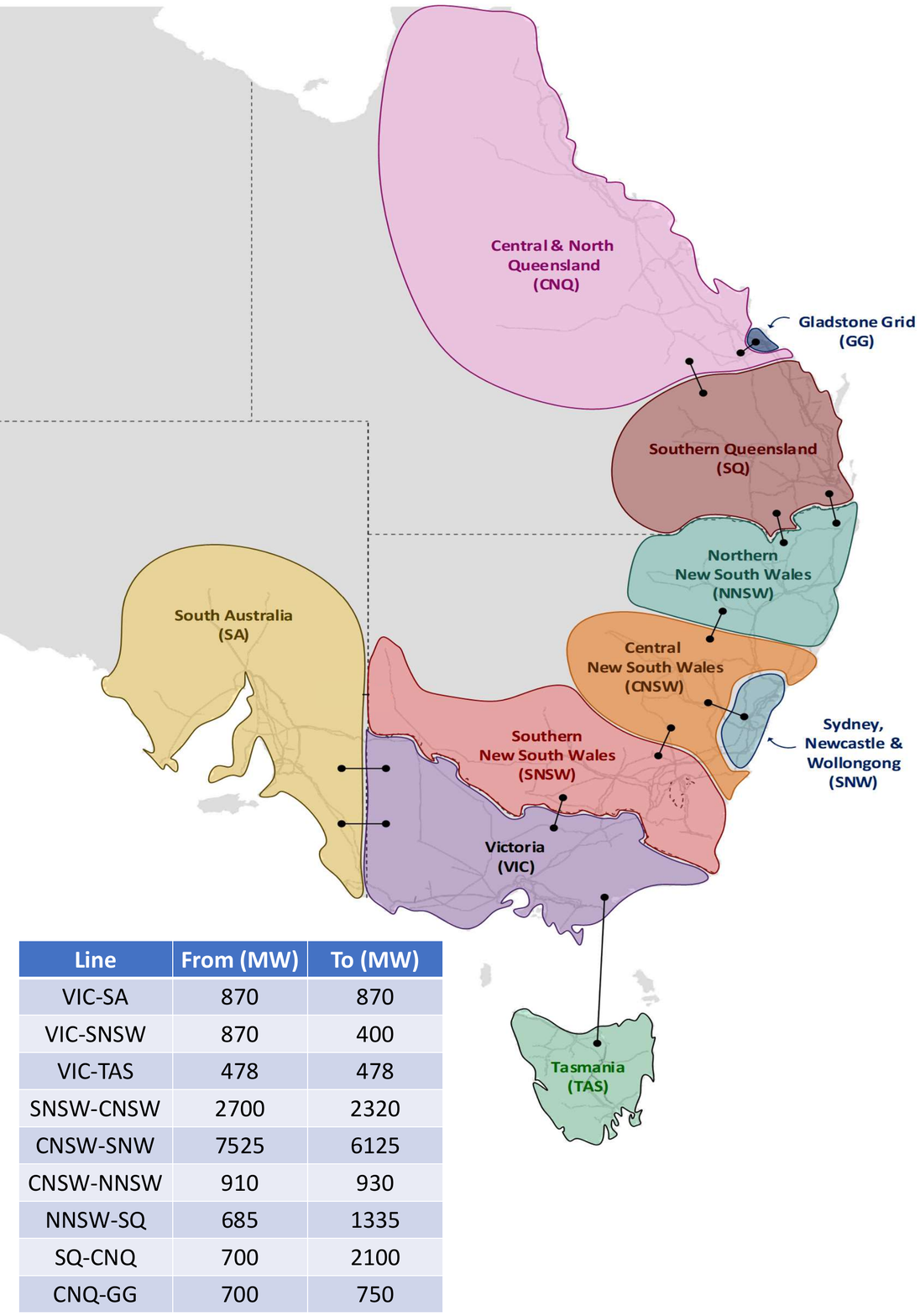}
	\caption{Modelled network topology of the National Electricity Market with ten local regions. Black Lines with adjoining dots represent transmission links between regions. Transmission network capacities between regions are indicated in the adjoining table.  Adapted from [35]}
	\label{fig:topology_NEM}
\end{figure}
\section{Extreme year scenarios used for Case Study}  \label{sec:App_B}
\begin{table} [H]%[thpb]
	\caption{Description of extreme year scenarios for case study}
	\renewcommand\arraystretch{1.3}
	\begin{center}
		\begin{tabular}{lll}
			\hline
			\# & Scenario & Description \\
			\hline
			1. & Extreme demand \& islanding - VIC/SA & Demand increased by 20\% over peak representative day.\\ 
			& & Availability for VIC-SA links constrained by 90\% \\
			2. & Extreme demand \& islanding - TAS & Demand increased by 40\% over peak representative day. \\ & & TAS-VIC link unavailable \% \\
			3. & Extreme demand \& islanding - QLD & Demand increased by 30\% over peak representative day.\\ & & Lines to northern QLD (SQ-CNQ and CNQ-GG) unavailable\\
			4. & High temperature thermal unavailability & Demand increased by 10\% over peak representative day.\\ & & Thermal generation availability across all regions reduced by 40\% \\
			5. & Renewables `dunkelflaute' & Demand increased by 10\% over peak representative day.\\ & & Variable renewable generation availability across all regions reduced by 80\% \\
			6. & Drought & Hydro inflows across all regions reduced by 20\% over year \\	
			\hline \hline
		\end{tabular}
	\end{center}
	\label{tab:extreme_scenarios_description}
\end{table}